\documentclass{chjaa}                  
\def \th {\thinspace}

\def\approxgt{\mathrel{\hbox{\rlap{\lower.55ex \hbox {$\sim$}} \kern-.3em \raise.4ex \hbox{$>$}}}}
\def\lesssim{\mathrel{\hbox{\rlap{\lower.55ex \hbox {$\sim$}} \kern-.3em \raise.4ex \hbox{$<$}}}}
\def\approxlt{\mathrel{\hbox{\rlap{\lower.55ex \hbox {$\sim$}} \kern-.3em \raise.4ex \hbox{$<$}}}}

\def \sun {\hbox {$\odot$}}

\usepackage{graphicx,times}             

\begin{document}

   \title{The periodic bursters XB\th 1323-619 and GS\th 1826-24: longterm evolution 
of the nuclear burning r\'egime and comparison with theory
}

   \volnopage{Vol.0 (200x) No.0, 000--000}      
   \setcounter{page}{1}                         

   \author{M. Ba\l uci\'nska-Church
      \inst{1,2}\mailto{}
   \and D. Reed
      \inst{1}
   \and M. J. Church
      \inst{1,2}
      }
   \institute{School of Physics and Astronomy, University of Birmingham, Birmingham B15 2TT, UK\\
             \email{mbc@star.sr.bham.ac.uk}
        \and
             Astronomical Observatory, Jagiellonian University, ul. Orla 171,
             30-244, Cracow, Poland\\
          }
   \date{Received~~2007 month day; accepted~~2007~~month day}

\abstract{The majority of X-ray burst sources do not display a burst rate that increases with
luminosity as expected, but this is seen in the two clocked bursters XB\th 1323-619 and GS\th 1826-24.
We present a detailed investigation of these two sources which in the case of the first source,
spans 18 years. Based on measurements of the burst rate, X-ray luminosity, the $\alpha$-parameter
and the two time constants generally present in the burst decays, we demonstrate the importance of
the {\it rp} nuclear burning process. A detailed comparison with theory shows that although the burst
rate in each source agrees well with the theoretical value, there is a difference of more than a factor
of 5 in the burst rate at a given luminosity between the sources. We show that the main reason for
this is that the two sources have substantially different emitting areas on the neutron star in non-burst 
emission, a factor often neglected. Variation of this area may explain the inverse relation of
burst rate with luminosity in the majority of burst sources.
   }

   \authorrunning{M. Ba\l uci\'nska-Church, D. Reed \& M. J. Church}            
   \titlerunning{Comparison of the X-ray burst rate with theory}                   

   \maketitle

%
\section{Introduction}           
\label{sect:intro}

X-ray bursting has been known for many years (Grindlay et al. 1976) as a phenomenon taking place in many low mass 
X-ray binaries (LMXB) of luminosities less than 10$^{38}$ erg s$^{-1}$ consisting of a rapid rise in intensity by 
a factor of $\sim$20, followed by an exponential decay over about 50 seconds and repeating on a timescale of hours. 
Bursting is usually not seen in higher luminosity sources forming the Z-track class which exhibit intensity 
increases as flaring over much longer timescales. Thus if we divide LMXB by luminosity into Atoll or Z-track 
sources, X-ray bursting is generally observed in the Atoll class. Bursting appears unrelated to inclination angle 
since it is seen not only in Atoll sources not displaying orbital-related behaviour but also in dipping sources
which are seen at high inclination. X-ray bursting has been extensively studied (Lewin et al. 1995; Strohmayer 
\& Bildsten 2006) and it was realized at an early stage to consist of unstable nuclear burning. Measured 
values of the $\alpha$-parameter, defined as the ratio of the energy released in steady mass accretion to the integrated 
energy of the burst, agreed well with values expected for unstable burning. Thus X-ray bursting was recognized
as explosive burning of recently accreted material on the surface of the neutron star (Woosley \& Taam 1976),
the material building up on the neutron star over a period of hours.

In spite of this, understanding of X-ray bursting is relatively poor. On the above basis the rate of X-ray bursting
would be expected to be generally stable when the luminosity of a sources is stable since then the mass accretion rate 
$\dot M$ is stable, but bursting rate is rather erratic in most sources. Even worse, the rate of bursting should increase
when source luminosity increases, as less time is required to accumulate the necessary mass and achieve the required
plasma density and temperature on the surface of the neutron star for unstable burning, but many sources display the
opposite of this (van Paradijs et al. 1988).

In two sources, which may be called the ``clocked bursters'', bursting is close to being exactly periodic and the burst 
rate increases with luminosity: GS\th 1826-24 (the original clocked burster) and XB\th 1323-619 which we have studied
over an extended period as one of the $\sim$10 dipping LMXB. In the present work, we make detailed comparisons of these 
two well-behaved sources with each other, and with the theory of unstable burning, to identify the reasons why the 
sources differ so much from each other, and to try to understand their behaviour in terms of theory. This would 
facilitate the understanding of the majority of burst sources that are not well-behaved. 
 

\begin{figure*}[!ht]                                                           
\begin{center}
\includegraphics[width=40mm,height=140mm,angle=270]{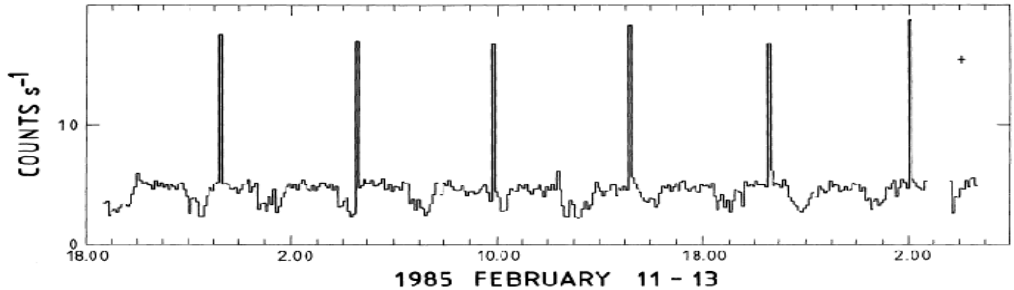}
\caption{{\it Exosat} lightcurve of XB\th 1323-619. Regular bursting takes place every 5.33 hr whereas X-ray
dipping occurs at the orbital period of 2.93 hr.}
\label{}
\end{center}
\end{figure*}

\section{Results of Analysis of XB\th 1323-619}
\subsection{Evolution of Burst Rate and Luminosity}

An example of regular bursting from the {\it Exosat} observation of XB\th 1323-619 is shown in Fig. 1
(Parmar et al. 1989). We have observed this source subsequently using {\it ASCA},
{\it BeppoSAX} (Ba\l uci\'nska-Church
et al. 1999), {\it Rossi-XTE} (Barnard et al. 2001) and {\it XMM} (Church et al. 2005). We show here a compilation
of results from these observations covering the period 1985 to the present. In the case of the {\it Exosat} observation
we have analysed the archival data to derive the burst decay time constants not previously obtained as
described below. In all of these observations
it was clear that the bursting remained regular; however the rate of bursting clearly increased. In all cases
the occurrence of bursting was consistent with a regular rate so that in sections of data without bursting,
the bursting would have occurred in data gaps due to Earth occultation and South Atlantic Anomaly passage.
The increase of burst rate was due to the increase in mass accretion rate shown by the systematically
increasing X-ray luminosity (see below and Fig. 3). In the case of the {\it XMM} observation, the burst rate
appeared less regular, and there was more variation in the burst height than previously seen.

\begin{figure*}[!h]                                                           
\begin{center}
\includegraphics[width=70mm,height=140mm,angle=270]{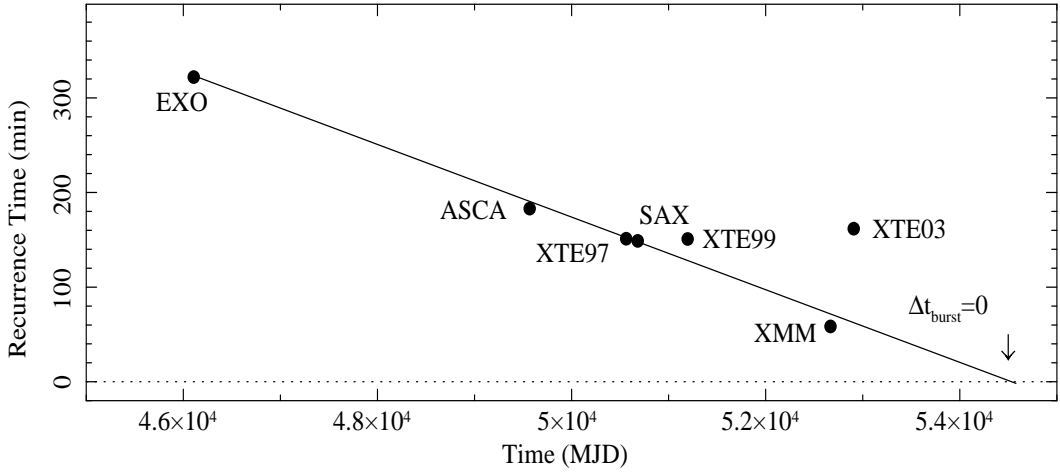}
\caption{Evolution of bursting in XB\th 1323-619 as a function of time: 1985 - 2003.}
\label{}
\end{center}
\end{figure*}

In Fig. 2 we show the mean time between bursts $\Delta$t as a function of time measured in Modified Julian Days 
for all of the observations
with X-ray telescopes since 1985, including 3 observations with {\it RXTE}. The last observation shown is the
2003 observation with {\it RXTE}. With the exception of this last observation there is a remarkable linearity
of $\Delta$t with MJD such that, if continued, $\Delta$t would become zero on January 11,
2008. In reality, $\Delta$t could not actually become zero, although a very fast burst rate may be possible.
In fact, the pattern of behaviour changes before the 2003 {\it RXTE} observation as the point for this observation
departs significantly from the linear relation of $\Delta$t versus time.

The 1 - 10 keV X-ray luminosity (non-burst) was increasing non-linearly during this period of 1985 - 2003, as shown in Fig.
3 (left panel), which also clearly shows the sharp drop in luminosity in the 2003 {\it RXTE} observation. 
However, the dependence of burst rate (1/$\Delta$t) on luminosity was linear, as shown in the right panel,
and this linear dependence confirms that XB\th 1323-619 behaves as expected from theory.
Moreover, when the luminosity decreased in 2003, the point for the 2003 {\it RXTE} observation 
continued to follow the same linear relation. 

\begin{figure*}[!ht]
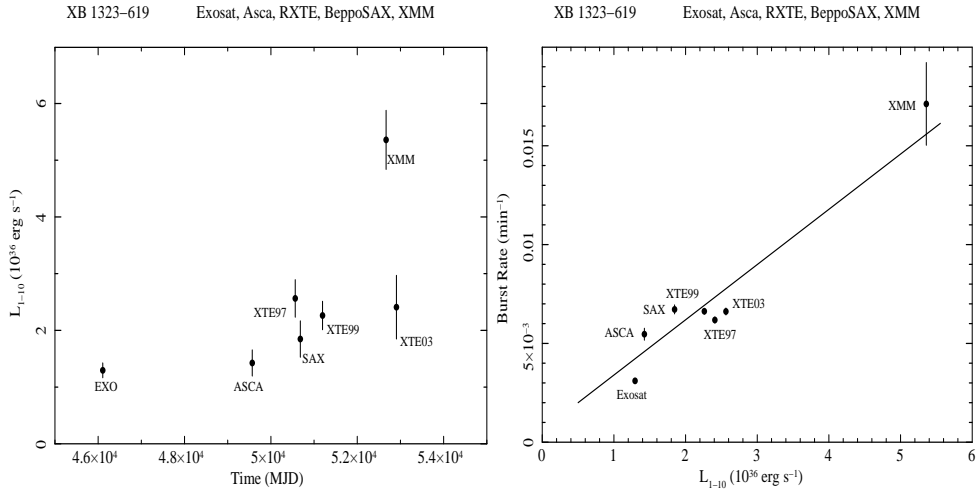
                                                           
\begin{center}
\includegraphics[width=64mm,height=64mm,angle=270]{balucinska_2007_revised_fig3a}        
\includegraphics[width=64mm,height=64mm,angle=270]{balucinska_2007_revised_fig3b}       
\caption{Left: evolution of X-ray luminosity with time; right: the linear dependence of burst rate
on luminosity.}
\label{}
\end{center}
\end{figure*}

\subsection{Evolution of Nuclear Burning R\'egime and Burst Profile}

In Fig. 4 we show our measurements of the $\alpha$-parameter, defined as the ratio of the non-burst fluence
to the burst fluence. In this figure, the bursts are shown equally spaced as a time sequence ignoring the large time
gaps between observations. The burst fluence was obtained by spectral fitting, for example, in the {\it RXTE} data
dividing the burst into 2-second long spectra and fitting each of these with a blackbody model so as to obtain the
flux in each segment, and adding these. In some cases, this was not possible because of the
limited number of counts in each burst. In this figure, we include a point for the single burst in the 1984
{\it Exosat} observation, which did not allow plotting in the previous figures as the burst rate was unknown.
It can be seen that until the {\it XMM} observation, there was
a constant $\alpha$ of $\sim$45. A value of 30 would show H burning and values of 100 more would indicate He burning
based on the known nuclear energy release per reaction, thus the results show mixed H and He burning. In {\it XMM}
a much higher $\alpha$ $\sim$100 - 150 was seen showing an increased contribution of He burning.

\begin{figure*}
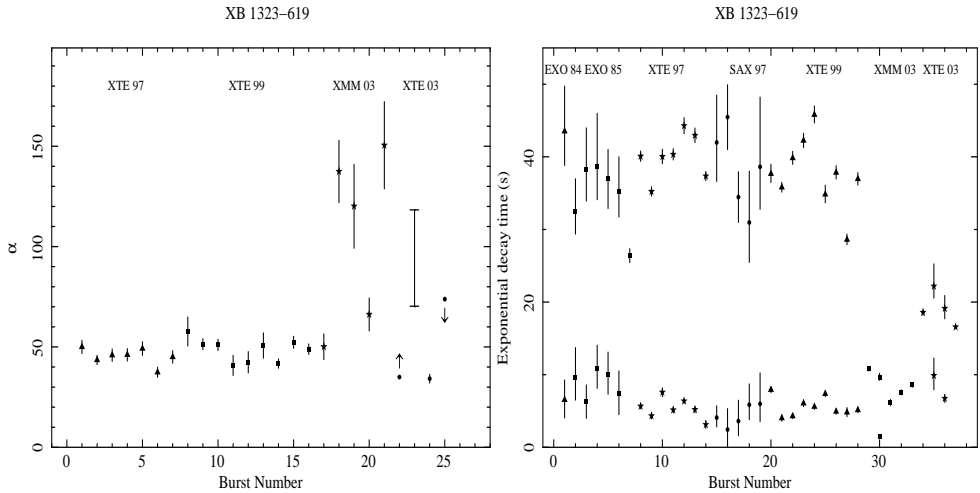
                                                             
\begin{center}
\includegraphics[width=64mm,height=64mm,angle=270]{balucinska_2007_revised_fig4a}                     
\includegraphics[width=64mm,height=64mm,angle=270]{balucinska_2007_revised_fig4b}                     
\caption{Left: Variation of the $\alpha$-parameter showing the strong change in the nature of nuclear burning; 
right: time constants of the burst decays: $\tau_1$ for the prompt decay, and $\tau_2$ for the later slow decay
of the burst (see text).}
\end{center}
\end{figure*}

For all of the observations, we also examined the exponential burst decays, fitting these to obtain the decay
time constants as shown in Fig. 5. Generally, each burst consisted of a normal fast decay with $\tau_1$ $\sim$5 s, 
followed by a much slower second decay, i.e. a long tail to each burst with $\tau_2$ $\sim$40 s, indicative of 
the rp burning process. A pre-cursor is formed in the initial fast burning but the further burning is
rate limited at so-called waiting points in the reaction chain. 
The rp-process takes place when the stable burning between bursts causes higher temperatures. Firstly, the CNO cycle
becomes the hot CNO cycle at $2\times 10^8$ K, since proton capture by $^{13}$N becomes faster than $\beta$-decay.
At $4\times 10^8$ K, production of $^{17}$F, $^{18}$Ne and $^{18}$F occurs, till at $8\times 10^8$ K, $^{21}$Na 
and $^{19}$Ne form break-out products for the cycle and these are the seeds for successive proton capture in the
rp-process forming heavy elements. In most of the historical observations, the long tail in each burst exists.
However, following the change of behaviour of bursting that was seen in the {\it XMM} observation, the long tail was not
seen and the subsequent reduction in $\tau_2$ (Fig. 4) indicates less rp-burning, a reason for which may be the
more complete burning of H to He inbetween bursts. This is consistent with the increase of $\alpha$ seen with {\it XMM}
as expected if the length of each burst is reduced so that the burst fluence decreases. 

\section{Comparison of XB\th 1323-619 and GS\th 1826-24}

During the present work we also carried out a reanalysis of the three {\it XTE} observations of
GS\th 1826-24 made in 1998, 2000 and 2002, analysis having previously been made by Galloway et
al. (2004). We found complete agreement with their work and in addition were able to obtain bolometric
X-ray luminosities in the wide band 0.1 - 200 keV as used below. In Fig. 5 we show the measured
burst rates in both sources as a function of X-ray luminosity $L$. In XB\th 1323-619 the 0.1 - 200 keV
flux and luminosity were also obtained from our best-fit spectral models in each observation.
Both sources behave as expected by simple theory, the burst rate increasing linearly with $L$.

\begin{figure*}[!h]
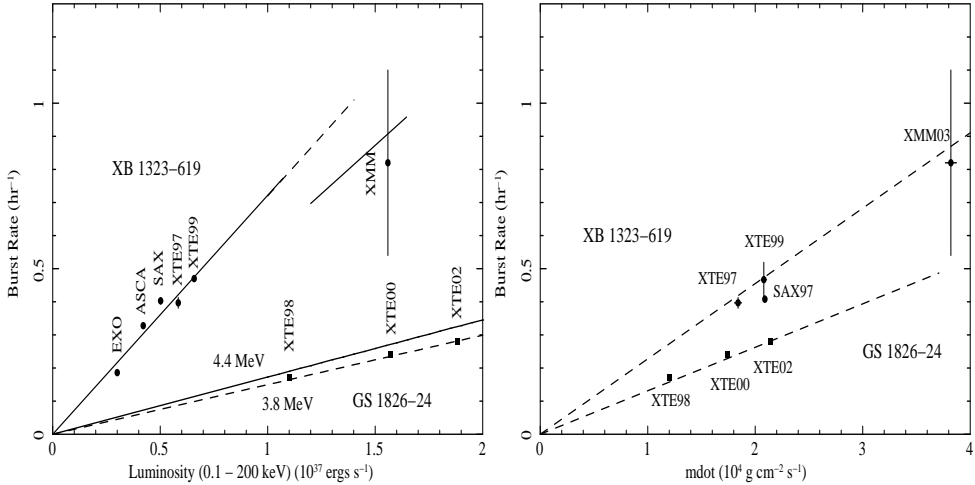
                                                            
\begin{center}
\includegraphics[width=64mm,height=64mm,angle=270]{balucinska_2007_revised_fig5a}  
\includegraphics[width=64mm,height=64mm,angle=270]{balucinska_2007_revised_fig5b}  
\caption{Comparison with theory. Left: burst rate as a function of luminosity showing the major difference in
burst rate between the two sources; right: burst rate re-plotted as a function of $\dot m$ showing the much
more consistent behaviour of the two sources.}
\label{}
\end{center}
\end{figure*}

\vfill\eject
The luminosity $L$ depends on mass accretion rate $\dot M$:

\[L = {GM\over R}\cdot {\Delta M\over \Delta t}\;\;\; {\rm so}\; {\rm that} \;\;\; \Delta t = {GM \Delta M\over RL}\] 

\noindent
where $\Delta t$ is the burst recurrence time and $\Delta M$ is the so-called ignition mass that accumulates
between bursts.
The ignition mass is obtained from the measured burst fluence $F$ and the theoretical energy release per
gram of burning matter $E$ via $\Delta M$ = $F/E$, where $E$ is 4.4 MeV per nucleon for the rp-process
(Fujimoto et al. 1987). In the case of XB\th 1323-619, $\Delta M$ is found to be $2.68\times 10^{20}$ g
leading to a theoretical rate of bursting: $(\Delta t)^{-1}$ = $0.719\,L_{37}$ hr$^{-1}$ for $M$ = 1.4$\,$M$_{\sun}$
and $R$ = 10 km. In Fig. 5 we plot
this theoretical relation which was obtained using data from the 1997 and 1999 {\it RXTE} observations.
It is clear that all of the data points lie well on this line, confirming that the nature of the nuclear burning
did not change in any of the observations before that with {\it XMM}. In {\it XMM}, the measured fluence changed
substantially so the theoretical line was re-calculated and plotted separately. In all cases, however, there was
good agreement with theory. In the case of GS\th 1826-24, Galloway et al. (2004) obtained the value of $E$ by
fitting the data as shown in Fig. 5. We show two theoretical lines: one for $E$ = 3.8 MeV per nucleon (dashed) as
obtained by Galloway et al., and also the line for 4.4 MeV per nucleon, which does not fit so well.

Thus in each source there is good agreement with theory derived on known energy releases and the measured
burst fluences in each case. However, the slopes of the linear dependences on $L$ differ by the large
factor of 5.5 showing a major differences between the two clocked bursters, and it is this difference
we aim to understand. The right panel of Fig. 5 shows the same results, but plotted as a function
not of luminosity  proportional to mass accretion rate $\dot M$, but of $\dot m$, the mass accretion rate
per unit X-ray emitting area of the neutron star, on the basis that this is the more important parameter
in the theory of stable and unstable nuclear burning (e.g. Bildsten 1998). It can immediately be seen
that the difference between the two sources is substantially reduced to the 40\% level. This is discussed
in more detail below.

\section{Discussion}

To obtain $\dot m$ as measured during a burst, we needed the blackbody emitting area, and this was
simply derived using the blackbody radius $R_{\rm BB}$ obtained at the peak of the burst by spectral
fitting, giving the area $A$ = $4\, \pi\,R_{\rm BB}^2$ and thus $\dot m$ = $\dot M / A$. Fig. 5 (right)
shows that although GS\th 1826-24 was at all times substantially more luminous than XB\th 1323-619,
the sources were more equal in terms of $\dot m$. The values of $\dot m$ of about $2\times 10^4$ 
g cm$^{-2}$ s$^{-1}$ lie in the r\'egime of nuclear burning where He burns unstably in a mixed H/He
environment (Fujimoto et al. 1981; Fushiki \& Lamb 1987; Bildsten 1998). We note that we could have 
assumed the whole neutron star was bursting but in this case the nuclear burning would be He burning
in a He environment - not as seen. 
Or we could possibly use the non-burst $R_{\rm BB}$ from spectral
fitting typically 1 km in which case $\dot m$ would be $2\times 10^5$ g cm$^{-2}$ s$^{-1}$, i.e.
in the stable region of nuclear burning - clearly incorrect, showing that the approach of measuring
$R_{\rm BB}$ at the peak of bursts was correct.

In our previous {\it ASCA} survey of LMXB, we investigated the relative luminosities of the neutron star
blackbody emission and that of the dominant Comptonized emission of the ADC, in view of the small
contribution of the blackbody in most LMXB when simple theory suggests it should be $\sim$50\%.
The results revealed a simple geometric relation that the height of the emitter on the neutron star
was equal to the height of the inner accretion disk obtained from the measured luminosity using
standard disk theory (Church \& Ba\l uci\'nska-Church 2001) valid over
3 decades of source luminosity. On the basis of this we are able to calculate the half-height of
the emitter $h$ of the non-burst emission in the two sources, which is difficult to measure by spectral
fitting in these weak sources, and compare this with the values derived from spectral fitting in bursts:

\vskip 2 mm
XB\th 1323-619: $h$ = 0.10 km,    burst $h$ = 1.4 km

GS\th 1826-24: $h$ = 0.75 km,     burst $h$ = 3.8 km

\vskip 2mm
It is clear that the smaller non-burst $h$ leads to a burst involving less area on the neutron star.
Clearly also, a spreading of the unstable burning takes place giving a large increase in area
compared with non-burst emission, and the larger the initial area, the more the area after spreading.

\section{Conclusions}

We have shown that marked differences in the burst rate occur between XB\th 1323-619 and GS\th 1826-24.
and have shown that the difference is mostly due to the very different emitting areas of non-burst emission
on the neutron star, a factor that has not generally been allowed for in discussion of X-ray bursting. If
we view the burst rate as a function of $\dot m$, the accretion rate per unit emitting area, we find
that the sources follow relations that agree within 40\%, so that both can be explained to this accuracy
by the same model. It is expected that the non-burst emitting area will be an important factor in
the other burst sources and in future work we will test the hypothesis that in sources where the burst
rate decreases with $L$, this behaviour can be explained in terms of the non-burst area increasing with $L$
so that $\dot m$ decreases.

\begin{acknowledgements}
This work was supported by the UK Particle Physics and Astronomy Research Council (PPARC) 
and the Polish Committee for Scientific Research grant no. KBN-1528/P03/2003/25. 
\end{acknowledgements}

\end{document}